\documentclass[twocolumn,prb,showpacs,preprintnumbers,amsmath,amssymb]{revtex4-1}
\usepackage{graphicx}% Include figure files
\usepackage{dcolumn}% Align table columns on decimal point
\usepackage{bm}% bold math
\usepackage{verbatim}

\begin{document}

%\preprint{APS/123-QED}

\title{Structural and $T_c$ inhomogeneities inherent to doping in La$_{2-x}$Sr$_x$CuO$_4$ superconductors and their effects on the precursor diamagnetism}

%Fluctuation diamagnetism around the superconducting transition in Tl$_2$Ba$_2$Ca$_2$Cu$_3$O$_{10}$
\author{Jes\'us Mosqueira}
\author{Luc\'ia Cabo}%
\author{F\'elix Vidal}

\affiliation{LBTS, Departamento de F\'isica da Materia Condensada, Universidade de Santiago de Compostela, Santiago de Compostela, E-15782 Spain}

\date{\today}% It is always \today, today,

\begin{abstract}
The inhomogeneities inherent to the random distribution of Sr dopants in La$_{2-x}$Sr$_x$CuO$_4$ superconductors are probed by measuring the x-ray diffraction linewidths and the Meissner transition widths, and then consistently explained on the grounds of a simple model in which the local Sr content is calculated by averaging over distances close to the in-plane electronic mean free path. By taking into account these intrinsic bulk inhomogeneities with long characteristic lengths (much larger than the superconducting coherence length amplitudes), the precursor diamagnetism measured above $T_c$, a fingerprint of the superconducting transition own nature, is then explained for all doping levels in terms of the conventional Gaussian-Ginzburg-Landau approach for layered superconductors. These results also suggest that the electronic inhomogeneities observed in the normal state by using surface probes overestimate the ones in the bulk.
\end{abstract}

\maketitle

\section{Introduction}

It is now well established that the presence of dopants in high-$T_c$ cuprate superconductors (HTSC), which so-deeply affect the behaviour of these materials by changing their carrier densities, may also cause structural, electronic, and critical temperature inhomogeneities, with different spatial distributions and characteristic lengths.\cite{balatsky} How these \textit{intrinsic} inhomogeneities inherent to doping affect the HTSC has now become a central open question of their physics,\cite{balatsky,lee} whose interest is also enhanced by the existence of a wide variety of doped materials which may be also deeply affected by inhomogeneities associated with doping.\cite{dagotto} However, in spite of their relevance, aspects as important as the spatial distribution of these inhomogeneities or their characteristic lengths are still not well settled. This last is crucial, in particular to understand how they affect the superconducting transition, at present another open and much debated issue also related to the controversy about the nature of the pseudogap in underdoped cuprates:\cite{balatsky,lee,timusk} These lengths could be of the order of the in-plane superconducting coherence length amplitude, $\xi_{ab}(0)$ (typically 2 to 4 nm in La$_{2-x}$Sr$_x$CuO$_4$), as is the case of the nanoscale electronic disorder observed on the surface of different HTSC by using scanning tunneling microscopy and spectroscopy (STM/S).\cite{balatsky,lee,mcelroy,gomes} In this case, these intrinsic inhomogeneities would deeply affect the own nature of their superconducting transition. In contrast, if they have in the bulk long characteristic lengths, much larger than $\xi_{ab}(0)$, they will just round the critical behaviour of any observable measured around $T_c$.\cite{reviewinhomo}

The present debate involving the entanglement between intrinsic inhomogeneities at different lengths scales, the pseudogap in the normal state and the superconducting fluctuations, is  particularly well illustrated by the diamagnetism observed above the superconducting transition: Earlier magnetization measurements in an underdoped cuprate (La$_{1.9}$Sr$_{0.1}$CuO$_4$) already suggested that the precursor diamagnetism measured in the bulk is conventional, i.e., that not too close to the transition it could be described in terms of the conventional Gaussian-Ginzburg-Landau (GGL) approach for layered compounds.\cite{PRLcarlos,mosqueira99} The anomalies observed in some cases, in particular under low magnetic fields, could be easily explained in terms of  $T_c$-inhomogeneities with long characteristic lengths [much larger than $\xi_{ab}(0)$], which do not directly affect the superconducting transition own nature.\cite{reviewinhomo,nuestros} Measurements of the in-plane paraconductivity\cite{paraconductivity} and of the heat capacity around $T_c$\cite{heatcapacity} in these underdoped cuprates, together with analyses of the effects of $T_c$-inhomogeneities with long characteristic lengths and of the background choice,\cite{reviewinhomo,ramallo00,maza91} fully agree with these conclusions. In particular, they confirm that above a reduced temperature of the order of $\varepsilon_c=0.5$ the superconducting fluctuation effects vanish, as predicted by taking into account the limits imposed by the uncertainty principle to the shrinkage of the superconducting wavefunction.\cite{cutoff} 

In the last years, however, different authors proposed that the anomalies observed in the precursor diamagnetism, mainly in measurements under low magnetic field amplitudes, provide compelling evidence that the superconducting transition of underdoped cuprates is unconventional, i.e., non describable in terms of the GGL approach for homogeneous layered superconductors.\cite{timusk,carretta,sewer,demello,wang,kresin,alexandrov,oganesyan,anderson,podolsky,salem,yuli} Nevertheless, these authors propose very different origins for such anomalous precursor diamagnetism, including the presence of superconducting phase fluctuations up to the pseudogap temperature or the existence in the bulk of intrinsic inhomogeneities with short characteristic lengths, as those observed by using surface probes.\cite{balatsky,lee,mcelroy,gomes} These different proposals enhance the interest of a thorough study of the inhomogeneities associated with doping and of their effects on the precursor diamagnetism.

In this paper, the $T_c$-inhomogeneities inherent to the random distribution of Sr dopants in La$_{2-x}$Sr$_x$CuO$_4$ superconductors are determined as a function of the doping level by applying to samples with high chemical quality two techniques that probe the bulk: powder x-ray diffraction (XRD) and low-field magnetization.\cite{pioneer} The observed XRD linewidths and $T_c$ dispersion are then explained on the grounds of a simple model in which the local Sr composition within the CuO$_2$ planes results from the average over distances  of the order of the in-plane electronic mean free path, of the order of 30 nm. This length is one order of magnitude larger than both $\xi_{ab}(0)$ and the characteristic length of the electronic disorder observed on the surface of different HTSC by using STM/S.\cite{balatsky,lee,mcelroy,gomes} The long characteristic length of the $T_c$-inhomogeneities inherent to doping is then fully confirmed by measurements of the precursor diamagnetism: The effects of these $T_c$-inhomogeneities may be quenched by just applying a magnetic field of moderate amplitude (of the order of 10\% the upper critical field amplitude), which shifts the average $T_c(H)$. Then, the remaining rounding effects on the magnetization above $T_c(H)$ may be explained at a quantitative level in terms of the conventional GGL approach for layered superconductors.

\section{Experimental}

\subsection{Preparation of the samples and characterization of their structural disorder}

For our study, we have chosen the La$_{2-x}$Sr$_x$CuO$_4$ system because it allows to cover the entire range of doping by varying the Sr content. Another central experimental aspect of our work here is to separate the chemical inhomogeneities inherent to doping from those extrinsic associated with a deficient growth process. So, to minimize these last we have decided to use powder samples because the Sr distribution may be very efficiently improved by successive processes of reaction and subsequent grinding. In addition, their small grain size ($4\pm 2\;\mu$m)  minimizes the presence of extrinsic structural inhomogeneities as compared to the typical mm-sized single crystals needed to perform the magnetization experiments. Finally, the difficulties associated with the random orientation of the individual grains and finite-size effects may be easily overcome.

\begin{figure}[b]
\includegraphics[scale=.45]{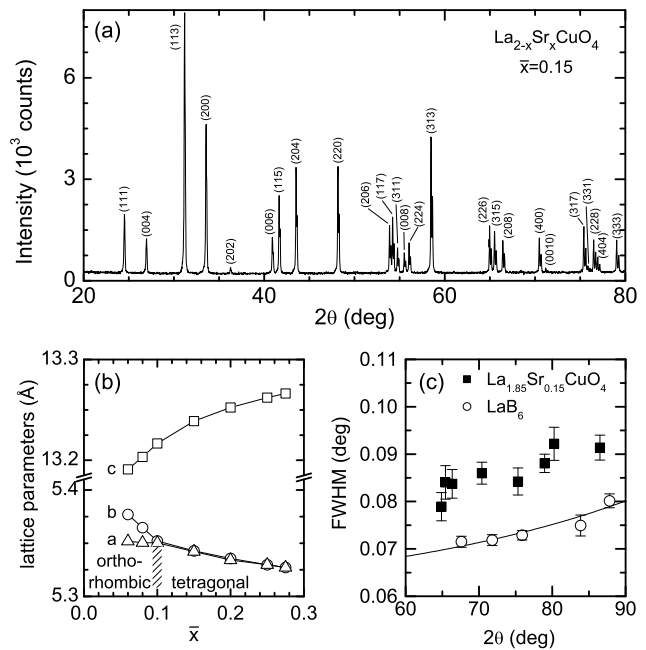}
\caption{a) Example of x-ray diffraction pattern, corresponding to the sample with $\bar x=0.15$ after 10 G-R steps. b) Resulting $\bar x$ dependence of the lattice parameters. c) Example of the XRD linewidths, corresponding to the $\bar x=0.15$ sample after 10 G-R steps, compared with the instrumental linewidths as determined from measurements in a reference LaB$_6$ powder sample (the line is a fit of a Caglioti function).}
\end{figure}

The polycrystalline La$_{2-x}$Sr$_x$CuO$_4$ samples with 0.06$\leq x\leq$0.275 were prepared by reacting in air at 930$^\circ$C for 20 h stoichiometric proportions of thoroughly mixed powders of La$_2$O$_3$, SrCO$_3$ and CuO (99.99\% purity). The resulting polycrystalline samples were ground and reacted again up to ten times. A small amount of each composition was taken after every two grind-react \mbox{(G-R)} steps in order to monitor the sample homogeneity. XRD measurements, performed with a Philips diffractometer equipped with a Cu anode and a Cu-K$_\alpha$ graphite monocromator, already excluded the presence of impurity phases after the second G-R step. 
An example of XRD pattern (corresponding to $\bar x=0.15$) is presented in Fig.~1(a). The resulting dependence of the lattice parameters on the average $x$ value ($\bar x$) is shown in Fig.~1(b), where the tetragonal-orthorhombic transition at $\bar x\approx0.10$ is clearly seen. When analyzed in detail, the diffraction peaks present a width significantly larger than the instrumental one. As an example, in Fig.~1(c) we present a comparison between the full-width at half-maximum (FWHM) of some diffraction peaks for the $\bar x=0.15$ sample after 10 G-R steps, with the instrumental FWHM. These FWHM values were obtained by fitting to the experimental peak profiles a pseudo-Voigt function [Fig.~2(a)]. It is worth mentioning that the linewidths decrease with the successive G-R steps: in the example of Fig.~2(a), after 2 G-R steps the Cu-$K_{\alpha_1}$ and $K_{\alpha_2}$ components of the (008) and (400) lines overlap, but after 10 steps their width is reduced and both components are resolved. However, as clearly shown in Fig.~2(b), no further reduction is observed above 8 G-R steps. The line broadening in excess of the instrumental one ($\Delta\theta$) may be then considered as \textit{intrinsic} and may be attributed to spatial variations of the lattice parameters ($\Delta a$, $\Delta b$ and $\Delta c$) around their average values ($\bar{a}$, $\bar{b}$ and $\bar{c}$), caused by the \textit{random} distribution of Sr dopants. 
%
%Let us note that the observation of the peak broadening already suggests that the characteristic length of these lattice distortions should not be significantly smaller than the \textit{transverse} x-ray correlation length (of the order of 100 nm).
%
%Let us note that the characteristic length of these lattice distortions must be much larger than the lattice parameters ($\sim10^{-10}$ m) although much smaller than the grains' size ($\sim10^{-6}$ m).

\begin{figure}[b]
\includegraphics[scale=.55]{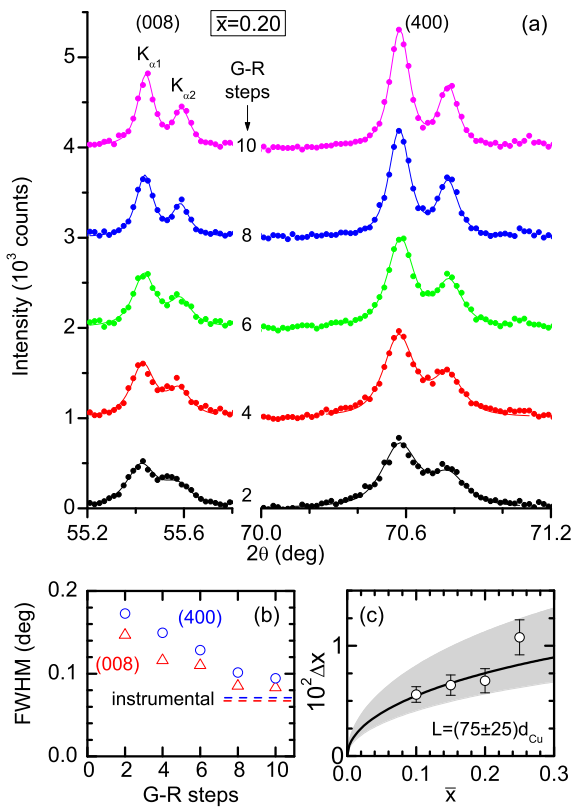}
\caption{(color online) a) Examples (for $\bar x$=0.20) of the XRD linewidths reduction with the successive G-R steps (different data sets are vertically displaced for clarity). The peaks' FWHM were obtained by fitting a pseudo-Voigt function (solid lines) taking into account the presence of both Cu-$K_{\alpha 1}$ and $K_{\alpha 2}$ wavelengths. As shown in (b), these FWHM saturate to a value significantly larger than the instrumental one after 8 G-R steps. c) $\bar x$ dependence of the corresponding \textit{intrinsic} $\Delta x$. The line corresponds to Eq.~(\ref{deltax}) evaluated with $L=75d_{\rm Cu}$ (the shaded area represents the uncertainty in this value).}
\end{figure}

\begin{figure}[t]
\includegraphics[scale=.55]{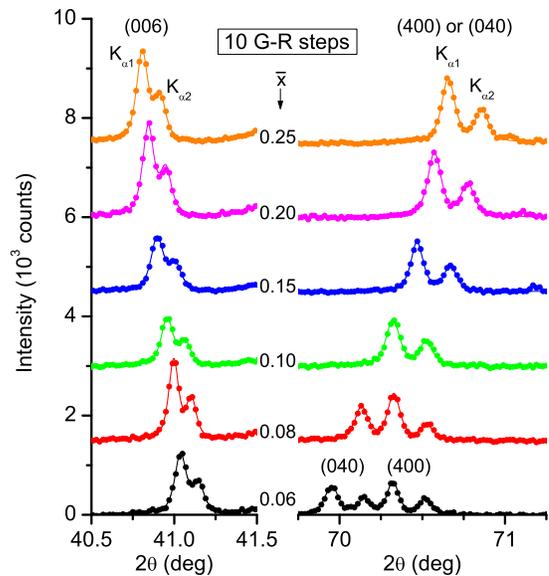}
\caption{(color online) Left panel: Example of the overlapping of the Cu-K$_{\alpha_1}$ and K$_{\alpha_2}$ diffraction lines at low diffraction angles (note that different datasets are vertically displaced). Right panel: Example of how the tetragonal-orthorhombic transition for $\bar x<0.10$ splits diffraction lines with nonzero $h$ and $k$ indexes, making difficult the linewidths determination.}
\end{figure}

For a general orthorhombic structure, the linewidth $\Delta\theta_{hkl}$ (corresponding to a diffraction line with indexes $h$, $k$ and $l$) is related to $\Delta a$, $\Delta b$ and $\Delta c$ through 
\begin{equation}
\sin\bar{\theta}_{hkl}\cos\bar{\theta}_{hkl}\Delta\theta_{hkl}\!=\!\frac{\lambda^2}{4}\!\left|\frac{h^2}{\bar{a}^3}\Delta a+\frac{k^2}{\bar{b}^3}\Delta b+\frac{l^2}{\bar{c}^3}\Delta c\right|,
\label{bragg}
\end{equation}
where $\lambda$ is the x-rays wavelength. Taking into account the dependence of the lattice parameters on the average $x$-value, Eq.~(\ref{bragg}) may be re-written as 
\begin{equation}
\Delta x=\frac{4\sin\bar{\theta}_{hkl}\cos\bar{\theta}_{hkl}}{\lambda^2\left|\frac{h^2}{\bar{a}^3}\frac{\partial \bar{a}}{\partial \bar x}+\frac{k^2}{\bar{b}^3}\frac{\partial \bar{b}}{\partial \bar x}+\frac{l^2}{\bar{c}^3}\frac{\partial \bar{c}}{\partial \bar x}\right|}\Delta\theta_{hkl},
\end{equation}
which allows to obtain the $x$ variation about $\bar x$ from each linewidth. In Fig.~2(c), we present the $\bar{x}$ dependence of $\Delta x$, as results from the analysis of lines within $60^\circ$~$\stackrel{<}{_\sim}$~$2\theta$~$\stackrel{<}{_\sim}$~$90^\circ$, in the samples after G-R steps. We discarded lower-$2\theta$ lines because of the overlapping of the $K_{\alpha_1}$ and $K_{\alpha_2}$ components (see Fig.~3, left panel), and higher-$2\theta$ lines because of their lower intensity. We also excluded samples with $\bar x<0.10$ because the tetragonal-orthorhombic transition splits the lines with nonzero $h$ and $k$ indexes, introducing a large uncertainty in the $\Delta\theta$ determination (see Fig.~3, right panel). In the range studied $\Delta x\sim0.01$, in agreement with previous estimations,\cite{pioneer} and it slightly increases with $\bar x$.

\begin{figure}[b]
\includegraphics[scale=.5]{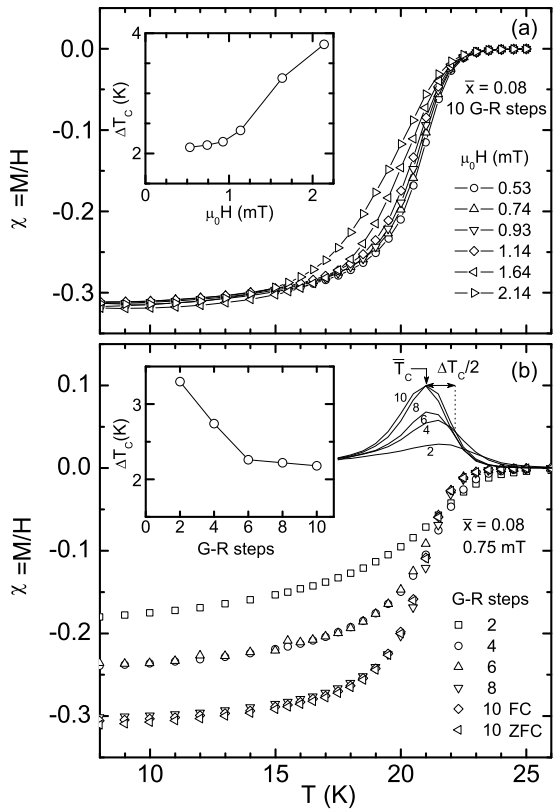}
\caption{a) Example (for the $\bar x=0.08$ sample after 10 \mbox{G-R} steps) of the as-measured $\chi^{FC}(T)$ under different applied magnetic fields. As shown in the inset, when \mbox{$\mu_0H<1$ mT} the transition width is almost unaffected by the use of a finite magnetic field. b) Example (for $\bar x=0.08$) of the dependence of the low-field $\chi^{FC}(T)$ on the number of G-R steps. The procedure used to obtain $\overline{T}_c$ and $\Delta T_c$ from $d\chi^{FC}/dT$ (solid lines) is also indicated. Inset: $\Delta T_c$ dependence on the number of G-R steps.}
\label{chivst}
\end{figure}

\begin{figure}[t]
\includegraphics[scale=.5]{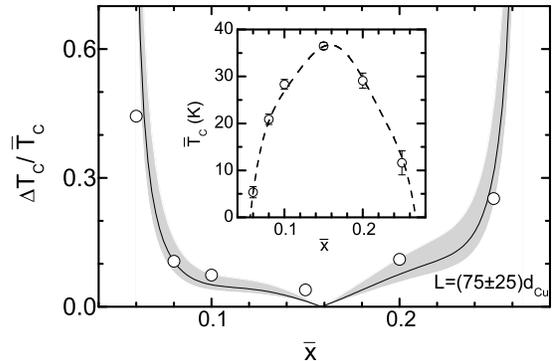}
\caption{$\bar x$ dependence of $\Delta T_c/\overline T_c$ and of $\overline T_c$ (inset), as obtained from the low-field $\chi^{FC}(T)$ measurements. The dashed line is a polynomial fit to data in the literature. The solid line corresponds to Eq.~(\ref{deltax}) evaluated with $L=75d_{\rm Cu}$ (the shaded area represents the uncertainty in this value).}
\label{uplot}
\end{figure}

\begin{figure*}[t]
\includegraphics[scale=.8]{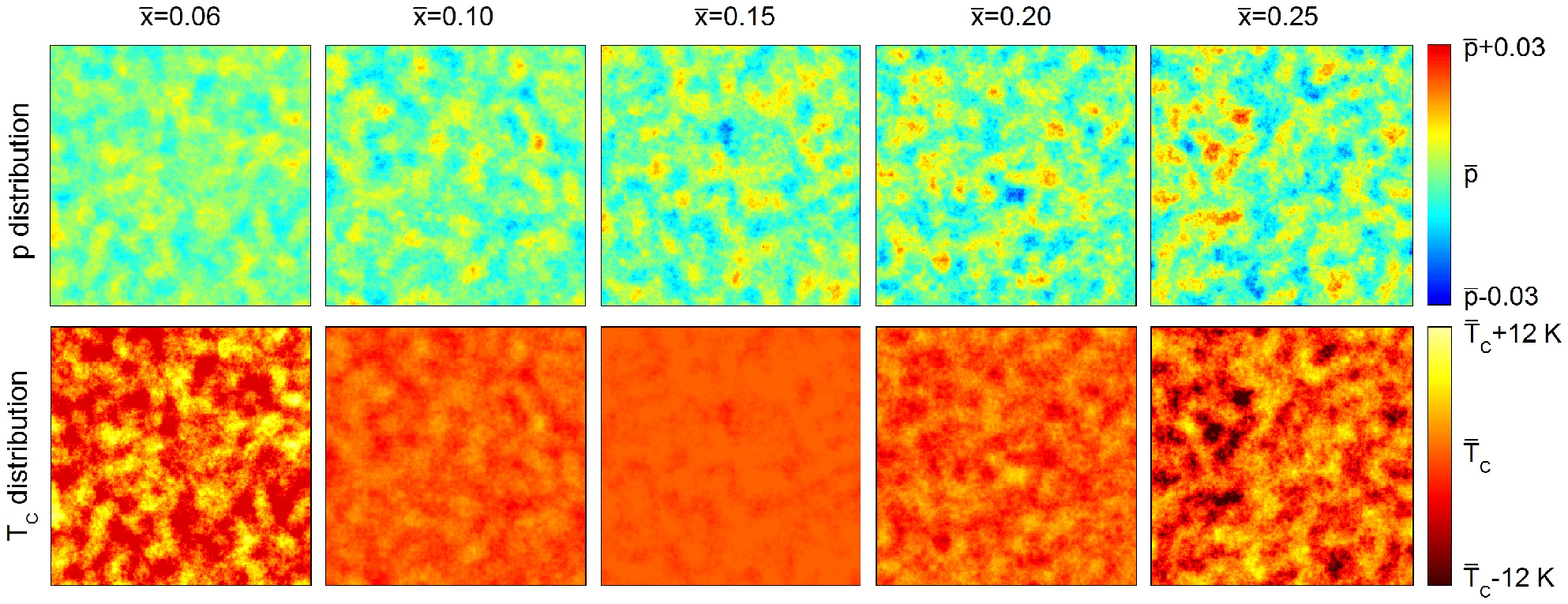}
\caption{(color online) Upper row: $x$ (or $p$) distributions in a $10^3\times10^3d_{\rm Cu}^2$ area within a CuO$_2$ plane for some representative $\bar x$ values, as calculated in terms of the model described in the text with $L=75d_{\rm Cu}$. The corresponding $T_c$ distributions are shown in the lower row.}
\label{panels}
\end{figure*}

\subsection{$T_c$ inhomogeneities associated with the random distribution of dopants, as probed by low-field magnetic susceptibility}

Due to the strong $T_c$ dependence on $\bar x$ in these materials,\cite{radaelli} the intrinsic compositional inhomogeneities measured above could lead to the existence of $T_c$ variations (characterized by a width $\Delta T_c$) around the average $\overline{T}_c$. Here we probe these intrinsic $T_c$ inhomogeneities through the temperature dependence of the low-field magnetic susceptibility measured under field-cooled (FC) conditions, $\chi^{FC}(T)$, by using a commercial DC magnetometer (Quantum Design, model PPMS). An example corresponding to $\bar x=0.08$ is shown in Fig.~\ref{chivst}. The magnetic field used in these measurements was always below 1 mT which, as shown in Fig.~\ref{chivst}(a), ensures that the contribution to $\Delta T_c$ associated with the transit through the mixed state is negligible. $\Delta T_c$ was obtained as twice the high-temperature half-width at half-maximum of the $d\chi^{FC}/dT$ versus $T$ curve. In this way, we elude the extrinsic rounding associated with the competition between the grains size and the magnetic penetration length (which is appreciable mainly below $\overline{T}_c$). Also, as in this region $|\chi^{FC}|\ll1$, demagnetization effects were neglected. As may be clearly seen in Fig.~\ref{chivst}(b), $\Delta T_c$ saturates after 8 G-R steps. This result, common to all compounds studied, is in agreement with the above XRD results, confirming the presence of an \textit{intrinsically} inhomogeneous charge distribution in the CuO$_2$ planes associated to the random distribution of dopants. The $\bar x$ dependence of $\Delta T_c/\overline{T}_c$ for the samples after 10 G-R steps is presented in Fig.~\ref{uplot}. As expected from the characteristic $\overline{T}_c(\bar x)$ dependence (inset of Fig.~\ref{uplot}) $\Delta T_c/\overline{T}_c$ is larger in the most underdoped and overdoped compounds.

\section{A model for the inhomogeneities associated to the random distribution of dopants: lenght scale for the hole concentration inhomogeneities}

Our results on $\Delta T_c(\bar x)$ may be accounted for by just assuming that the random distribution of Sr ions induces spatial variations of the hole concentration within a CuO$_2$ plane, with a characteristic length $L$. The local hole concentration $p$ (holes/CuO$_2$) (or, equivalently, the local $x$ value) may be calculated by averaging over a ``control'' circle of radius $L$ within the CuO$_2$ plane, centered on that site, 
\begin{equation}
p=x=N_{\rm Sr}\frac{d^2_{\rm Cu}}{\pi L^2}.
\label{px}
\end{equation}
Here $N_{\rm Sr}$ is the number of Sr ions within that circle, and $d_{\rm Cu}\approx0.38$ nm is the Cu-Cu shortest distance. The smaller the $L$ value, the larger will be the statistical differences between the local and the average hole concentrations.\cite{singer05} In the upper row of Fig.~\ref{panels} we present the $p$ (or $x$) distribution in a $10^3\times10^3d^2_{\rm Cu}$ area within a CuO$_2$ plane, as calculated on the basis of this model for several $\bar{x}$ values representative of the over-, optimal- and under- doped regimes. These maps were calculated by taking into account that there are two (La/Sr)O layers associated with each CuO$_2$ plane, and that the probability for a La ion to be substituted by a Sr ion is $\bar{x}/2$. Also, we used $L=75d_{\rm Cu}\approx28$ nm, which is the value leading to the best agreement with the above experimental results (see below). As expected, $p$ varies over lengths of the order of $L$, and the spread $\Delta p$ around the corresponding average value $\bar p$ increases with doping. As $L$ is much larger than the in-plane superconducting coherence length amplitude [$\xi_{ab}(0)\approx2-4$ nm, see also below], and by virtue of the $T_c(\bar x)$ relation, this $p$ (or $x$) spatial distribution leads to the $T_c$ spatial distribution shown in the lower row of Fig.~\ref{panels}. As may be clearly seen, $T_c$ also varies on the scale of $L$, but now the spread in $T_c$ values is larger in compounds where $d\overline T_c/d\bar{x}$ is larger, i.e., highly underdoped or overdoped. 

We may now obtain expressions for $\Delta p$ and $\Delta T_c$, by noting that the probability that $N_{\rm Sr}$ out of the $N_L$ sites for the Sr ions within the control circle are occupied is given by the binomial distribution
\begin{equation}
\delta(N_{\rm Sr})=\left(
\begin{array}{c}
N_L\\
N_{\rm Sr}
\end{array}
\right)\left(\frac{\bar{x}}{2}\right)^{N_{\rm Sr}}\left(1-\frac{\bar{x}}{2}\right)^{N_L-N_{\rm Sr}}.
\label{binomial}
\end{equation}
Here $N_L$ may be approximated by the integer closest to $2\pi L^2/d_{\rm Cu}^2$, the factor 2 accounting for the two (La,Sr)O layers associated with each CuO$_2$ layer. In the limit of large $N_L$ (more precisely, if $N_L\bar{x}/2\gg5$, which for $L/d_{\rm Cu}=75$ is well satisfied), Eq.~(\ref{binomial}) may be approximated by a normal distribution centered in $\overline N_{\rm Sr}=\bar x\pi L^2/d_{\rm Cu}^2$ and with FWHM $\Delta N_{\rm Sr}\approx1.66\sqrt{N_L\bar x(1-\bar x/2)}$. By combining this with Eq.~(\ref{px}), the corresponding hole density distribution is also normal, centered in $\bar p=\bar x$ and with FWHM
\begin{equation}
\Delta p=\Delta x\approx1.33\frac{d_{\rm Cu}}{L}\sqrt{\bar x\left(1-\frac{\bar x}{2}\right)}.
\label{deltax}
\end{equation}
In turn, as for the $L$ value used in the simulations $\Delta x\ll \bar x$, the corresponding $\Delta T_c$ may be approximated by
\begin{equation}
\Delta T_c(\bar x)\approx\Delta x(\bar x)\left|\frac{\partial \overline T_c}{\partial \bar x}\right|.
\label{deltaTc}
\end{equation}

The solid line in Fig.~\ref{uplot} is the best fit of Eq.~(\ref{deltaTc}) to the experimental data. The agreement is good, taking into account the experimental uncertainties, and it leads to $L=(75\pm25)d_{\rm Cu}\approx28\pm9$ nm. It is remarkable that Eq.~(\ref{deltax}) evaluated with the same $L$ value is in excellent agreement with the $\Delta x(\bar x)$ dependence deduced from x-ray diffraction [Fig.~2(c)], which suggests a strong correlation between the structural and electronic disorders. Also, this $L$ value is in reasonable agreement with a \textit{universal length scale} recently proposed for the charge density inhomogeneity in these compounds for temperatures around $T_c$ ($\sim30$ nm, see Fig.~2 in Ref.~\onlinecite{mihailovic05}). It is, however, larger than the length scale for the spatial variations in the local density of states as determined by STM (5-10 nm at 4 K).\cite{balatsky,kato05} 

\section{Fluctuation-diamagnetism above $\overline T_c$: a check of consistency for the length scale of the inhomogeneities inherent to doping}

One may also probe the characteristic length of the $T_c$ inhomogeneities associated with doping through measurements of the fluctuation-induced magnetization, $M_{fl}$, above $\overline T_c$. This observable is very sensitive to the domain size relative to the in-plane superconducting coherence length, $\xi_{ab}$:\cite{reviewinhomo} for reduced temperatures $\varepsilon\equiv\ln(T/\overline T_c)$ such that $\xi_{ab}(\varepsilon)>L$, $M_{fl}$ would have a zero-dimensional (0D) character, otherwise it would be two-dimensional (2D). By using the mean-field relation $\xi_{ab}(\varepsilon)=\xi_{ab}(0)\varepsilon^{-1/2}$, the 2D-0D crossover would happen for $\varepsilon\approx\xi_{ab}^2(0)/L^2$. In the case that $L\sim\xi_{ab}(0)$ (as suggested by STM) all the accessible $\varepsilon$-range would be in the 0D regime. 

\begin{figure}[b]
\includegraphics[scale=.5]{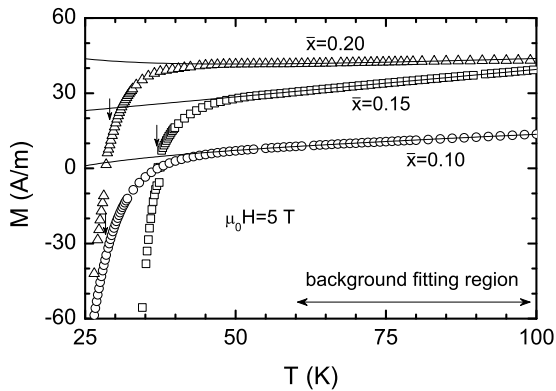}
\caption{Some examples of the as-measured $M(T)$ in the normal state. The effect of superconducting fluctuations is clearly seen as a rounding close to the corresponding $T_c$ (indicated by arrows). The normal-state background (solid lines) was determined by fitting a Curie-like function above 60 K. }
\label{backs}
\end{figure}

\begin{figure}[t]
\includegraphics[scale=.5]{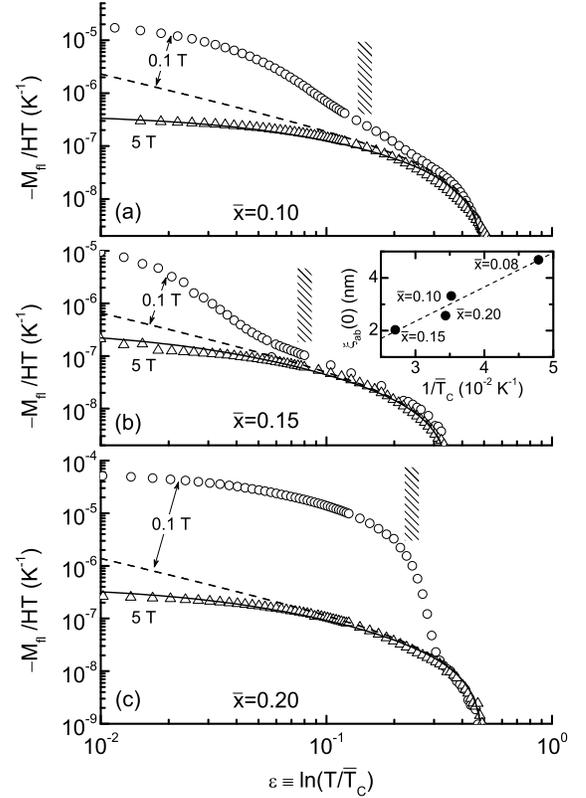}
\caption{Reduced-temperature dependence of $M_{fl}$ (over $HT$) for different doping levels. The dramatic effects of $T_c$ inhomogeneities inherent to doping manifested at low fields are quenched by the application of a 5 T magnetic field, allowing a direct comparison with the 2D-GGL theory (lines). Dashed bars are the expected onset of the $|M_{fl}|$ upturn due to these $T_c$ inhomogeneities at low fields. Inset: The resulting $\xi_{ab}(0)$ values are presented against $T_c^{-1}$. }
\label{mfl}
\end{figure}

Some examples of the $M(T)$ curves above $T_c$ for different doping levels (as measured with a Quantum Design SQUID magnetometer) are presented in Fig.~\ref{backs}. From these measurements $M_{fl}(T)$ was obtained by subtracting to $M(T)$ the normal state contribution, which in turn was determined by fitting a Curie-like function [$M_{\rm back}(T)=A+BT+C/T$, being $A$, $B$ and $C$ fitting constants] in a temperature interval well above $T_c$, where fluctuation effects are expected to be negligible. Note that whereas the background uncertainties may have a dramatic effect on the extracted $M_{fl}$ at high reduced temperatures ($\varepsilon>10^{-1}$), their influence is small for $\varepsilon\stackrel{<}{_\sim}10^{-1}$, the relevant region for our present analyses.
Some examples of the resulting $M_{fl}(\varepsilon)$ (for convenience over $HT$) for different doping levels are presented in Fig.~\ref{mfl}. As may be clearly seen, under a 0.1~T magnetic field [much smaller than $H_{C2}^\perp(0)$, the upper critical field for $H\perp ab$ extrapolated to $T = 0$~K], $|M_{fl}|$ presents an upturn as $\varepsilon\to0$. This behavior may be easily explained as a consequence of the $T_c$ distribution, and should occur below $\varepsilon_{\rm inh}\sim2\Delta T_c/\overline T_c$.\cite{nuestros} From the $\Delta T_c$ data in Fig.~\ref{uplot}, the resulting $\varepsilon_{\rm inh}$ (dashed bars in Fig.~\ref{mfl}) are in good agreement with the onset of the observed enhanced $|M_{fl}|$. In the presence of a finite magnetic field, the region affected by inhomogeneities, i.e., around $\overline T_c(H)$, is shifted to lower temperatures. This allows to compare the theoretical results for homogeneous materials with the experimental data in a wider temperature region above $\overline T_c(0)$. Quantitatively, it may be approximated $\varepsilon_{\rm inh}(H)\sim2\Delta T_c/\overline T_c-H/H_{C2}^\perp(0)$.\cite{nuestros} As $\mu_0H_{C2}^\perp(0)\sim20-80$ T in these materials,\cite{hc2} a 5 T magnetic field will allow to study $M_{fl}$ free from the effect of $T_c$ inhomogeneities down to $\varepsilon\sim0$ for $0.1\stackrel{<}{_\sim}\bar x\stackrel{<}{_\sim}0.2$.

In presence of a finite magnetic field perpendicular to the CuO$_2$ planes, $M_{fl}^\perp$ in the framework of the 2D Gaussian Ginzburg-Landau (GGL) approach is given by  \cite{carballeira}
\begin{eqnarray}
M_{fl}^\perp=-f\frac{k_BT}{\phi_0s}\left[-\frac{\varepsilon_c}{2h}\psi\left(\frac{h+\varepsilon_c}{2h}\right)-\ln\Gamma\left(\frac{h+\varepsilon}{2h}\right)\right.+\nonumber \\
+\left.\ln\Gamma\left(\frac{h+\varepsilon_c}{2h}\right)+\frac{\varepsilon}{2h}\psi\left(\frac{h+\varepsilon}{2h}\right)+\frac{\varepsilon_c-\varepsilon}{2h}\right].
\label{prange}
\end{eqnarray}
Here $\Gamma$ and $\psi$ are the gamma and digamma functions, $h\equiv H/H_{C2}^\perp(0)$ the reduced magnetic field, $s\approx6.6$~\r{A} the periodicity length of the superconducting layers, $f$ the effective superconducting fraction,\cite{meissner} $k_B$ the Boltzmann constant, $\phi_0$ the flux quantum, and $\varepsilon_c\approx0.5$ the total-energy cutoff constant.\cite{cutoff}
In the low magnetic-field limit ($h\ll\varepsilon$) Eq.~(\ref{prange}) may be approximated by
\begin{equation}
M_{fl}^\perp=-f\frac{k_BT}{6\phi_0s}h\left(\frac{1}{\varepsilon}-\frac{1}{\varepsilon_c}\right),
\label{Schmidt2D}
\end{equation}
which in absence of a cut-off ($\varepsilon_c\to\infty$) reduces to the well known Schmid result for 2D materials.\cite{schmid}
As in these materials the anisotropy factor, $\gamma$, is in the range 10-25,\cite{gamma} the fluctuation magnetization when $H\parallel ab$ (of the order of $\sim  M_{fl}^\perp/\gamma^2$, see e.g. Ref.~\onlinecite{mparallel}) is negligible. Then, $M_{fl}$ in a powder sample with the grains randomly oriented may be approximated by the angular average\cite{mosqueira99} 
\begin{equation}
\langle M_{fl}\rangle=\frac{1}{2}\int_0^{\pi}M_{fl}^\perp(H\cos\theta)\cos\theta \sin\theta d\theta,
\end{equation}
where $\theta$ is the angle between the $c$-axis and $H$.
As may be clearly seen in Fig.~\ref{mfl}, the comparison of the 2D-GGL theory (solid lines) with the 5~T data is excellent down to the lowest accessible reduced temperature ($\varepsilon\sim10^{-2}$), the resulting $\xi_{ab}(0)=[\phi_0/2\pi \mu_0H_{C2}^\perp(0)]^{1/2}$ being about 2-4 nm (inset in Fig.~\ref{mfl}) in agreement with recent estimates.\cite{hc2} The 2D nature of $M_{fl}$ is consistent with a length scale for the in-plane modulation of the hole-concentration  well above the nanometer scale, suggesting that the STM results\cite{balatsky,lee,timusk,mcelroy,gomes} could not be representative of the bulk. 

\section{Conclusions}

The results we have obtained in the prototypical La$_{2-x}$Sr$_x$CuO$_4$ system provide unambiguous answers to some of the questions we have stressed in the introduction: i) The unavoidable inhomogeneities when doping the cuprate superconductors have long characteristic lengths [much larger than $\xi_{ab}(0)$] and they are uniformly distributed in the bulk. ii) These intrinsic inhomogeneities may deeply affect the bulk magnetization measurements around $T_c$, but once they are taken into account the diamagnetic transition may be described for all doping regions in terms of the Ginzburg-Landau approach for layered superconductors, i.e., from a phenomenological point of view the intrinsic diamagnetic transition in the HTSC is \textit{conventional}. Additionally, our results do not support the presence of fluctuating vortices well above the superconducting transition, even in the pseudogap regime, and they also suggest that the different inhomogeneities observed by using surface probes overestimates the ones in the bulk. 

\section*{Acknowledgments}

Supported by the Spanish Ministerio de Educaci\'on y Ciencia (Grant No. FIS2007-63709), and the Xunta de Galicia (Grants No. 07TMT007304PR and 2006/XA049), in part with FEDER funds.


\begin{references}


\bibitem{balatsky}For reviews and references see, e.g., A.V.~Balatsky, I.~Vekhter, and Jian-Xin~Zhu, Rev. Mod. Phys. \textbf{78}, 373 (2006); \O.~Fischer, M.~Kugler, I.~Maggio-Aprile, Ch.~Berthod, and Ch.~Renner, Rev. Mod. Phys. \textbf{79}, 353 (2007).
  
\bibitem{lee}For a review and references see, e.g., P.A.~Lee, N.~Nagaosa, Xiao-Gang~Wen, Rev. Mod. Phys. \textbf{78}, 17 (2006); see also A.N.~Pasupathy, A.~Pushp, K.K.~Gomes, C.V.~Parker, Jinsheng~Wen, Zhijun~Xu, Genda~Gu, S.~Ono, Y.~Ando, and A.~Yazdani, Science \textbf{320}, 196 (2008), and references therein.

\bibitem{dagotto}See e.g., E. Dagotto, Science \textbf{309}, 257 (2005), and references therein.

\bibitem{timusk}For introductory reviews and references see e.g., T. Timusk, Physics World \textbf{18}, 31 (2005); M. Franz, Nature Physics \textbf{3}, 686 (2007); B.G. Levi, Phys. Today \textbf{60}, 17 (2007). 

\bibitem{mcelroy}K.~McElroy, J.A.~Jinho Lee, D.H.~Slezak, H.~Lee, H.~Eisaki, S.~Uchida, and J.C.~Davis, Science \textbf{309}, 1048 (2005).

\bibitem{gomes}K.K.~Gomes, A.N.~Pasupathy, A.~Pushp, S.~Ono, Y.~Ando, and A.~Yazdani, Nature \textbf{447}, 569 (2007).

\bibitem{reviewinhomo}For a review and references see, e.g., F.~Vidal, J.A.~Veira, J.~Maza, J.~Mosqueira and C.~Carballeira, in \textit{Material Science, Fundamental Properties and Future Electronic Applications of High-$T_c$ Superconductors}, NATO ASI Series, edited by S.L.~Dreschler and T.~Mishonov (Kluver- Dordrecht, Amsterdam, 2001), p. 289. See also, arXiv:cond-mat/0510467 (unpublished).

\bibitem{PRLcarlos}C.~Carballeira, J.~Mosqueira, A.~Recovlevschi and F.~Vidal, Phys. Rev. Lett. \textbf{84}, 3157 (2000).

\bibitem{mosqueira99}J.~Mosqueira, M.V.~Ramallo, A.~Revcoleschi, C.~Torr\'on, and F.~Vidal, Phys. Rev. B {\bf 59}, 4394 (1999). 

\bibitem{nuestros}L.~Cabo, F.~Soto, M.~Ruibal, J.~Mosqueira, and F.~Vidal, Phys. Rev. B {\bf 73}, 184520 (2006); L.~Cabo, J.~Mosqueira, and F.~Vidal, Phys. Rev. Lett. {\bf 98}, 119701 (2007). See also, J.~Mosqueira, L.~Cabo, and F.~Vidal, Phys. Rev. B \textbf{76}, 064521 (2007); J.~Mosqueira and F.~Vidal, Phys. Rev. B \textbf{77}, 052507 (2008).

\bibitem{paraconductivity}S.R.~Curr\'as, G.~Ferro, M.T.~Gonzalez, M.V.~Ramallo, M.~Ruibal, J.A.~Veira, P.~Wagner, and  F.~Vidal,  Phys. Rev. B \textbf{68}, 094501 (2003); S.H.~Naquib, J.R.~Cooper, J.L.~Tallon, R.S.~Islan, and R.A.~Chakalov, \textit{ibid}. \textbf{71}, 054502 (2005); T.~Aoki, Y.~Oikawa, C.~Kim, T.~Tamura, H.~Ozaki, and N.~Mori, Physica C \textbf{463-465}, 126 (2007). See also, B. Leridon, J. Vanacken, T. Wambecq, and V.V. Moshchalkov, Phys. Rev. B \textbf{76}, 012503 (2007).


\bibitem{heatcapacity}J.W.~Loram, J.L.~Tallon, and W.Y.~Liang, Phys. Rev. B \textbf{69}, 060502 (2004); J.W.~Loram and J.L.~Tallon, \textit{ibid.} \textbf{79}, 144514 (2009). Although these authors analyze the fluctuation effects on the heat capacity close to $T_c$ in terms of the 3D-XY approach for the full critical region, they also conclude the absence of bulk inhomogeneities with short characteristic lengths.

\bibitem{ramallo00}M.V. Ramallo and F.~Vidal, Phys. Rev. Lett. \textbf{85}, 3543 (2000); M.V.~Ramallo, C.~Carballeira, and F.~Vidal, Physica C \textbf{34}, 173(2000).

\bibitem{maza91}For earlier analyses of the entanglement between $T_c$-inhomogeneities with long characteristic lengths and uniformly distributed in the bulk and the superconducting fluctuations see, J.~Maza and F.~Vidal, Phys. Rev. B \textbf{43}, 10560 (1991). 

\bibitem{cutoff}F.~Vidal, C.~Carballeira, S.R.~Curr\'as, J.~Mosqueira, M.V.~Ramallo, J.A.~Veira and J.~Vi\~na, Europhys. Lett. {\bf 59}, 754 (2002).

\bibitem{carretta}P. Carretta, A. Lascialfari,  A. Rigamonti,A. Roso, and A. A. Varlamov, Phys. Rev. B 61, 12420(2000); A.A. Lascialfari,  A. Rigamonti, L. Romano, A. A. Varlamov, and I. Zuca, ibid. 68, 100505 (2003).

\bibitem{sewer}A. Sewer and H. Beck, Phys. Rev. B \textbf{64}, 014510 (2001). See also, S. Weyeneth, T. Schneider, and E. Giannini, \textit{ibid}. \textbf{79}, 214504 (2009), and references therein.

\bibitem{demello}E.V.L. de Mello, E.S. Caxeiro, and J.L. Gonz\'alez, Phys. Rev. B \textbf{67}, 024502 (2003); J.L. Gonz\'alez and E.V.L. de Mello, \textit{ibid}. \textbf{69}, 134510(2004).

\bibitem{wang}Y.  Wang, L. Li, and N. P. Ong,  Phys. Rev. B \textbf{73}, 024510 (2006), and references therein; N.P. Ong, Y. Wang, L. Lu, and M.J. Naughton,  Phys. Rev. Lett. \textbf{98}, 119720 (2007). 

\bibitem{kresin}V. Kresin,  Z. Ovchinnikov, Y.N. and S.A. Wolf,  Phys. Rep. \textbf{431}, 231 (2006), and references therein.

\bibitem{alexandrov}A.S. Alexandrov, Phys. Rev. Lett. \textbf{96}, 147003 (2006).

\bibitem{oganesyan}V. Oganesyan,  D.A. Huse, and S.L. Sondhi,   Phys. Rev. B \textbf{73}, 09450 (2006).

\bibitem{anderson}P.W. Anderson, Phys. Rev. Lett. \textbf{96}, 017001 (2006). See also arXiv:cond-mat/0504453(unpublished);  Nature Physics \textbf{3}, 160 (2007).

\bibitem{podolsky}D. Podolsky, S. Raghu, and A. Vishwanath, Phys. Rev. Lett. \textbf{99}, 1117004 (2007).

\bibitem{salem}S. Salem-Sugui, Jr., M.M. Doria, A.D. Alvarenga, V.N. Vieira, P.F. Fariñas, and J.P. Sinnecker, Phys. Rev. B \textbf{76}, 132502 (2007); S. Salem-Sugui, Jr., J. Mosqueira, and A. D. Alvarenga, Phys. Rev. B \textbf{80}, 094520 (2009).

\bibitem{yuli}O. Yuli, I. Asulin, O. Millo, and D. Orgad, Phys. Rev. Lett. \textbf{101}, 057005 (2008).

\bibitem{pioneer}A pioneering application of both techniques to study the inhomogeneities in LSCO compounds may be seen in J.B.~Torrance, Y.~Tokura, A.I.~Nazzal, A.~Bezinge, T.C.~Huang, and S.S.P.~Parkin, Phys. Rev. Lett. \textbf{61}, 1127 (1988); see also H.~Takagi, R.J.~Cava, M.~Marezio, B.~Batlogg, J.J.~Krajewski, W.F.~Peck, Jr., P.~Bordet, and D.E.~Cox, Phys. Rev. Lett. {\bf 68}, 3777 (1992).

\bibitem{radaelli}P.G.~Radaelli, D.G.~Hinks, A.W.~Mitchell, B.A.~Hunter, J.L.~Wagner, B.~Dabrowski, K.G.~Vandervoort, H.K.~Viswanathan, and J.D.~Jorgensen, Phys. Rev. B {\bf 49}, 4163 (1994).

\bibitem{singer05}A similar procedure has been used to account for the line broadening of the $^{17}$O NMR spectrum in the normal state of La$_{2-x}$Sr$_x$CuO$_4$ in P.M. Singer {\it et al.}, Phys. Rev. B {\bf 72}, 014537 (2005).

\bibitem{mihailovic05}D.~Mihailovic, Phys. Rev. Lett. {\bf 94}, 207001 (2005).

\bibitem{kato05}T. Kato, S. Okitsu, and H. Sakata, Phys. Rev. B {\bf 72}, 144518 (2005).

\bibitem{hc2}Y.~Wang and H-H.~Wen, Europhys. Lett. \textbf{81}, 57007 (2008).

\bibitem{carballeira}C.~Carballeira, J.~Mosqueira, A.~Revcoleschi, and F.~Vidal, Physica C \textbf{384}, 185 (2003).

\bibitem{meissner}See e.g., J.~Mosqueira, J.A.~Camp\'a, A.~Maignan, I.~Rasines, A.~Revcolevschi, C.~Torr\'on, J.A.~Veira, and F.~Vidal, Europhys. Lett. {\bf 42}, 461 (1998).

\bibitem{schmid}A.~Schmid, Phys. Rev. {\bf 180}, 527 (1969).

\bibitem{gamma}T.~Shibauchi, H.~Kitano, K.~Uchinokura, A.~Maeda, T.~Kimura, and K.~Kishio, Phys. Rev. Lett. {\bf 72}, 2263 (1994).

\bibitem{mparallel}Z.~Hao and J.~Clem, Phys. Rev. B {\bf 46}, 5853 (1992).



\end{references}
\end{document}